\documentclass[sigconf]{acmart}

\usepackage{algorithm2e}
\usepackage{amsmath,amsfonts}
\usepackage{algorithmic}
\usepackage{graphicx}
\usepackage{textcomp}
\usepackage{xcolor}
\usepackage{url}
\usepackage{subfig}

\newcommand{\eventually}[0]{\Diamond}
\newcommand{\globally}[0]{\Box}

\newcommand{\actionseq}[0]{\textbf{a}}
\newcommand{\until}[0]{\mathcal{U}}
\newcommand{\signal}[0]{\textbf{s}}
\newcommand{\Signal}[0]{\textbf{S}}

\newcommand{\weakerthan}[0]{\ll}

\newcommand{\strongerthan}[0]{\gg}

\newcommand\parameter{\textbf{p}}
\newcommand\valuation{\textbf{v}}

\newtheorem{problem}{Problem}[section]
\AtBeginDocument{%
  }

\setcopyright{acmlicensed}
\copyrightyear{2018}
\acmYear{2018}
\acmDOI{XXXXXXX.XXXXXXX}

\acmConference[Conference acronym 'XX]{Make sure to enter the correct
  conference title from your rights confirmation emai}{June 03--05,
  2018}{Woodstock, NY}
\acmISBN{978-1-4503-XXXX-X/18/06}

\copyrightyear{2024}
\acmYear{2024}
\setcopyright{rightsretained}
\acmConference[SEAMS '24]{19th International Symposium on Software Engineering for Adaptive and Self-Managing Systems}{April 15--16, 2024}{Lisbon, AA, Portugal}
\acmBooktitle{19th International Symposium on Software Engineering for Adaptive and Self-Managing Systems (SEAMS '24), April 15--16, 2024, Lisbon, AA, Portugal}
\acmDOI{10.1145/3643915.3644090}
\acmISBN{979-8-4007-0585-4/24/04}

\begin{document}

\title{
Integrating Graceful Degradation and Recovery through Requirement-driven Adaptation
}
\author{Simon Chu}
\email{cchu2@andrew.cmu.edu}
\affiliation{%
  \institution{Carnegie Mellon University}
  \city{Pittsburgh}
  \state{PA}
  \country{USA}
}

\author{Justin Koe}
\email{jkoe469@gmail.com}
\affiliation{%
  \institution{The Cooper Union}
  \city{New York}
  \state{NY}
  \country{USA}
}

\author{David Garlan}
\email{dg4d@andrew.cmu.edu}
\affiliation{%
  \institution{Carnegie Mellon University}
  \city{Pittsburgh}
  \state{PA}
  \country{USA}
}

\author{Eunsuk Kang}
\email{eunsukk@andrew.cmu.edu}
\affiliation{%
  \institution{Carnegie Mellon University}
  \city{Pittsburgh}
  \state{PA}
  \country{USA}
}


\begin{abstract}
    Cyber-physical systems (CPS) are subject to environmental uncertainties such as adverse operating conditions, malicious attacks, and hardware degradation. These uncertainties may lead to failures that put the system in a sub-optimal  or unsafe state. Systems that are resilient to such uncertainties rely on two types of operations: (1) \emph{graceful degradation}, for ensuring that the system maintains an acceptable level of safety during unexpected environmental conditions and (2) \emph{recovery}, to facilitate the resumption of normal system functions. Typically, mechanisms for degradation and recovery are developed independently from each other, and later integrated into a system, requiring the designer to develop an additional, ad-hoc logic for activating and coordinating between the two operations.

    In this paper, we propose a self-adaptation approach for improving system resiliency through automated triggering and coordination of graceful degradation and recovery. The key idea behind our approach is to treat degradation and recovery as \emph{requirement-driven} adaptation tasks: Degradation can be thought of as temporarily \emph{weakening} original (i.e., ideal) system requirements to be achieved by the system, and recovery as \emph{strengthening} the weakened requirements when the environment returns within an expected operating boundary. Furthermore, by treating weakening and strengthening as \emph{dual operations}, we argue that a single requirement-based adaptation method is sufficient to enable coordination between degradation and recovery. Given system requirements specified in \emph{signal temporal logic (STL)}, we propose a run-time adaptation framework that  performs degradation and recovery in response to environmental changes. We describe a prototype implementation of our framework and demonstrate the feasibility of the proposed approach using a case study in unmanned underwater vehicles.

\end{abstract}





\maketitle

\section{Introduction}


Cyber-physical systems (CPS) encompasses systems with both physical and software components, such as autonomous vehicles, unmanned aerial vehicles (UAVs), smart grids, and robots. These systems are subject to environmental uncertainties such as adverse operating conditions (e.g., severe weather for vehicles), malicious attacks, and hardware faults (e.g., damaged or flaky sensors).
Due to these uncertainties, CPS may encounter failures during their operation lifecycle. For example, an autonomous vehicle may deviate from the lane boundaries due to a distracted driver or bad road conditions, drones may enter unsafe airspace due to unexpected turbulences, and a city may encounter blackouts due to unexpectedly high demands on the electricity grid. 

Achieving resiliency against such failures typically involves two types of operation~\cite{CPSResilienceMetric}: (1) \emph{graceful degradation}, for ensuring that the system maintains its most critical safety functions during unexpected environmental scenarios and (2) \emph{recovery}, to enable the resumption of normal functions as the environment returns to its expected state. Typically, mechanisms for degradation and recovery are developed independently and later integrated into a single system. For instance, a designer of an automotive safety architecture may combine a degradation mechanism that uses secondary sensors (e.g., cameras) when primary ones fail (e.g., Lidar under inclement weather) and another mechanism for re-activating an optimal system function (e.g.,  self-driving mode) when the environmental conditions improve (e.g., Lidar providing accurate data again). There are two challenges with this approach: (i) The designer is responsible for developing and validating an application-specific logic that decides when degradation or recovery should be activated and (ii) as system requirements evolve over time, this logic may also need to be changed.

In this paper, we propose a self-adaptation approach for improving system resiliency through automated coordination of graceful degradation and recovery. The key idea behind our approach is to treat degradation and recovery as \emph{requirement-driven} adaptation tasks: Degradation can be thought of as temporarily \emph{weakening} an original system requirement to be achieved by the system, and recovery as \emph{strengthening} a previously weakened requirement when the environmental conditions improve. Furthermore, weakening and strengthening can be regarded as \emph{dual} operations: The former relaxes the set of system behaviors that are deemed acceptable, and the latter restricts it. Based on this idea, we propose a single, unified requirement-driven adaptation framework that is capable of automatically switching between degradation and recovery, depending on the changes that arise in the environment. Our proposed approach overcomes the above two challenges, by (i) removing the need to develop an application-specific logic for coordinating degradation and recovery, and (ii) allowing the designer to plug in a different requirement without modifying the underlying logic.

To concretize this approach, we propose a self-adaptation framework that takes system requirements specified in \emph{signal temporal logic (STL)}~\cite{stl}, a type of temporal logic that is particularly well-suited for specifying time-varying behaviors of CPS (e.g., ``The vehicle must never deviate outside the lane for more than 2 seconds''). We show how the weakening and strengthening operations can be formulated formally as the problem of relaxing and strengthening a given STL specification, respectively. In addition, it would be desirable to reduce the impact of degradation (i.e., apply minimal weakening necessary) and maximize the rate of recovery (i.e., strengthen the requirement as much as possible). To support such \emph{optimal} degradation and recovery, we also describe how the problem of generating \emph{minimal weakening} and \emph{maximal strengthening} can be encoded and solved as an instance of \emph{mixed-integer linear programming (MILP)}.

We have developed a prototype implementation of our proposed adaptation framework. To demonstrate its feasibility, we have applied the framework to a case study involving an \emph{unmanned underwater vehicle (UUV)}, where the system may encounter environmental uncertainties such as low water visibility and thruster failures while on a mission to inspect an underwater pipeline \cite{suave-case-study}.
We compare our approach against the state-of-the-art adaptation framework called TOMASys~\cite{meta-model}. Our experimental results are promising, showing that our approach can achieve a higher level of requirement satisfaction throughout the adaptation process while incurring a reasonable amount of overhead.

In summary, our main contributions are as follows:

\begin{enumerate}
    \item A theoretical framework that combines graceful degradation and recovery using incremental weakening and strengthening of STL-based requirements.
    

    \item A runtime architecture that performs weakening and strengthening to support system adaptation given changing environmental conditions.
   
    \item An approach for enabling optimal system behavior by finding a minimal weakening or maximum strengthening through MILP.
    
  
    \item An implementation of the proposed adaptation framework and a case study involving UUVs.
\end{enumerate}
\medskip

\section{Motivating Example}
\label{section:motivating-example}

\begin{figure}[h]
\centering
\includegraphics[width=0.4\textwidth]{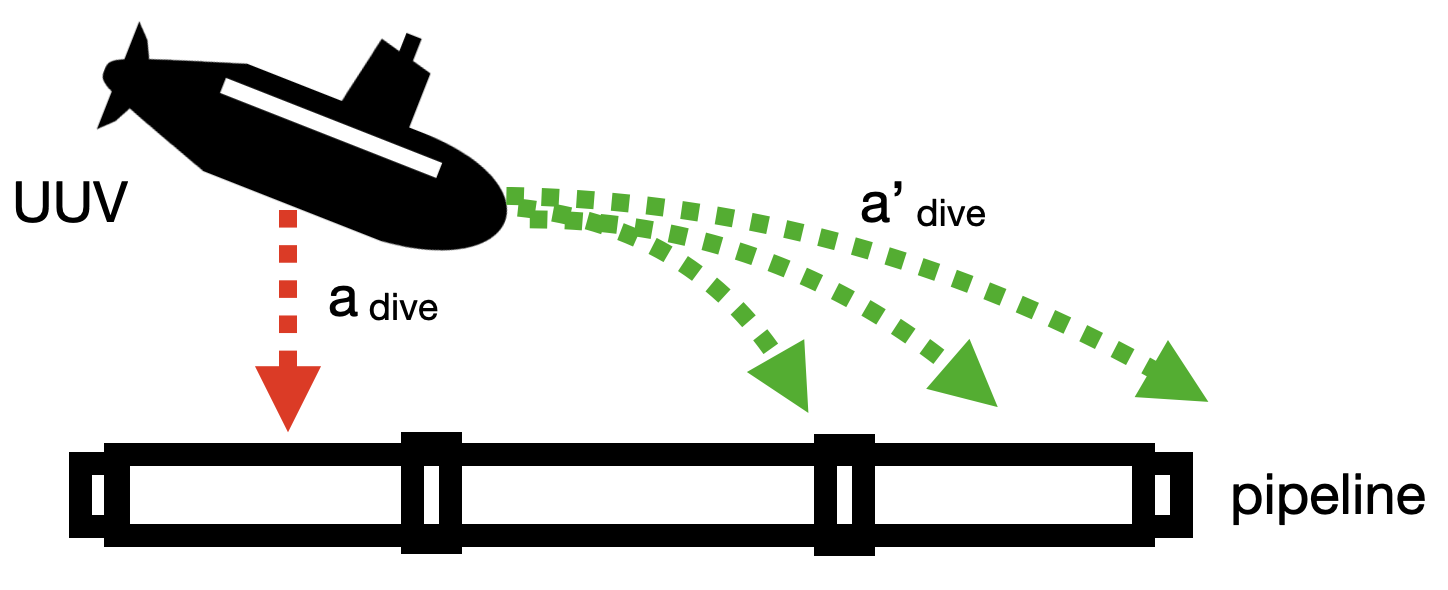}
\caption{Illustration of UUV inspecting underwater pipeline
}
\label{fig:motivating-example}
\end{figure}

Consider an UUV (illustrated in Figure \ref{fig:motivating-example}) that is on a mission to inspect underwater pipelines, inspired by the SUAVE exemplar system~\cite{suave-case-study}. The mission involves the UUV simultaneously following and inspecting a pipeline. There are two mission objectives that the system aims to achieve. First, it must maintain a clear line of sight with the underwater pipeline; if the visual contact is lost, the UUV must regain the  contact within the next few seconds. Second, the thruster in the UUV must continually provide enough thrust to allow the vehicle to complete the mission within given time $T$.

During the mission, the UUV is subject to two sources of uncertainties: ($U_1$) change in water visibility and ($U_2$) thruster failures. These uncertainties may result in the system failing to meet the mission objectives stated above. For example, $U_1$ may result in the UUV losing visual contact with the pipeline and unable to regain the visual contact in time.  $U_2$ may result in the engine being unable to provide enough thrust to complete the mission in time. In either scenario, the violation of the mission objectives may result in hardware damage and loss of the mission entirely.

\paragraph{Existing Approach}
To safely degrade the functionality of the UUV during an adverse environmental condition, the developer may first construct a monitor that raises a warning when water visibility falls below a certain threshold (e.g., 5 meters) and triggers a certain degradation action (e.g., reduce the speed of the system or remain stationary to prevent collision with surrounding obstacles). Separately, to support recovery, another monitor may be created to look for when the water visibility improves and perform appropriate recovery actions (e.g., dive deeper  to regain its visual contact). 

This approach, however, has some drawbacks. First, the developer is responsible for designing triggering conditions and response actions, and validating that these actions maintain a desirable level of system utility (e.g., safety); this requires considerable domain knowledge and manual effort on the developer's part. Second, when system requirements evolve, these conditions and response actions may also need to be changed. For example, suppose that the UUV is required to float back up to the surface instead of diving deeper when the visibility falls below a certain safe threshold. Supporting this new requirement would involve designing a new controller (specific to this requirement) that triggers degradation and recovery based on the changes in the environmental condition.

\paragraph{Proposed Approach}


To overcome these drawbacks, we take a \emph{requirements-driven} adaptation approach. Given a user-specified requirement, our approach automatically switches between degradation and recovery by weakening and strengthening the requirement, respectively, and adjusting the system behavior to adapt to the modified requirement. In addition, our approach can determine an \emph{optimal} amount of weakening or strengthening that is needed to degrade the system safely or bring it back to a normal operational state.

For example, suppose that the user-specified  requirements for the \emph{visibility} and \emph{thruster} features are as follows:
\begin{quote}
$R_{visibility}$: \emph{Every time the visual contact with the submarine is lost, regain the  contact within the next 5 seconds by diving deeper toward the pipeline.}\\
$R_{thruster}$: \emph{The thruster should provide 100 N of thrust to allow it to complete the mission on time.}
\end{quote}

During an adverse environmental event, each of the requirements can be weakened to adapt to the changing environment. For example, the \emph{visibility} requirement may be weakened by changing the time to regain contact from within the next 5 second to within the next 15 seconds in the case of severely low water visibility, as follows:
\begin{quote}
$R_{visibility}'$: \emph{Every time the visual contact with the submarine is lost, regain the contact within the next 15 seconds by diving deeper toward the pipeline.}
\end{quote}
This weakened requirement $R_{visibility}'$ allows the visibility monitor to either delay the action to regain visual contact or descend towards the pipeline at a slower rate to ensure the safety of the vehicle. The weakening of the requirement can, in turn, increase the range of control actions the system can select from, shown in Figure \ref{fig:motivating-example} (action space increases from $a_{dive}$ to $a_{dive}'$ after weakening). Then, once the visibility has improved significantly, the vehicle may want to resume normal operation by reducing the time it takes to regain contact, by strengthening $R'_{visibility}$ back to $R_{visibility}$.

Similarly, in the case of an engine failure, the \emph{thruster}  requirement can be weakened by changing the required thrust  from 100 N to some value less than 100 N, under the premise that it still provides an acceptable level of utility. One such possible weakening is as follows:
\begin{quote}
    $R_{thruster}'$: \emph{The thruster should provide 50 N of thrust to allow it to complete the mission on time.}
\end{quote}
This weakened requirement $R_{thruster}'$ allows certain thrusters to be turned off. Once the engine recovers (through a repair), the framework again identifies the best possible requirement for the \emph{thruster} feature, ensuring that the UUV completes the mission in the most timely manner---by strengthening $R'_{thruster}$ and allowing the system to turn thrusters back on.




\paragraph{Challenges}
Generally, given a requirement like $R_{visibility}$, there are numerous ways of weakening or strengthening this requirement (e.g., adjusting the time to regain visibility by different amounts). Weakening a requirement involves temporarily sacrificing a certain utility while strengthening leads to a regain of that utility. One challenge is how to systematically weaken the requirement no more than needed while strengthening the requirement to the maximum extent possible. Another challenge is given a weakened or strengthened requirement, how to reconfigure or adjust the behavior of the system to best fulfill the adjusted requirement. To tackle these challenges,  we will describe (1) how requirement weakening and strengthening can be formalized using \emph{parametric signal temporal logic (PSTL)}\cite{PSTL} and (2) how to perform optimal requirement adaptation by encoding it as a MILP problem.

\section{Preliminaries}

\paragraph{Signals}
In our approach, the behavior of CPS is modeled by real-valued continuous-time \emph{signals}.
Formally, a signal is a function $\signal{}:T\rightarrow D$ mapping from a time  domain, $T\subseteq \mathbb{R}_{\geq 0}$, to a tuple of $k$ real numbers, $D\subseteq \mathbb{R}^k$. The value of a signal $\signal{}(t) = (v_1,\dots, v_k)$ represents different state variables of the system at time $t$ (e.g., $v_1$ might represent the altitude of a drone).

\paragraph{Signal temporal logic (STL)}
STL extends linear temporal logic (LTL) \cite{ltl} for specifying the time-varying behavior of a system in terms of signals. The basic unit of a formula in STL is a signal predicate in the form of $f(\signal{}(t)) > 0$, where $f$ is a function from $D$ to $\mathbb{R}$; i.e., the predicate is true if and only if $f(\signal{}(t))$ is greater than zero. The syntax of an STL formula $\varphi$ is defined as:
\begin{equation*}
    \varphi := f(\signal{}(t)) > 0 \mid \neg \varphi \mid \varphi_1\wedge \varphi_2\mid \varphi_1 \mathcal{U}_{[a,b]}\varphi_2
\end{equation*}
where $a,b\in \mathbb{R}$  and $a<b$. The \emph{until} operator $\varphi_1 \mathcal{U}_{[a,b]}\varphi_2$ means that $\varphi_1$ must hold until $\varphi_2$ becomes true within a time interval $[a,b]$.
The until operator can be used to define two other important temporal operators: \emph{eventually} ($\Diamond_{[a,b]}\varphi := True\ \mathcal{U}_{[a,b]}\varphi$) and \emph{always} ($\Box_{[a,b]}\varphi := \neg\Diamond_{[a,b]}\neg \varphi$). 

\paragraph{Robustness}
Typically, the semantics of temporal logic such as LTL is based on a \emph{binary} notion of  satisfaction (i.e., formula $\varphi$ is either satisfied or violated by the system). Thanks to its signal-based nature, STL also supports a \emph{quantitative} notion of satisfaction, which allows reasoning about how ``close" or ``far'' the system is from satisfying or violating a property. This quantitative measure is  called the \emph{robustness} of satisfaction~\cite{FainekosP09}. 

Informally, the robustness of signal $\signal{}$ with respect to formula $\varphi$ at time $t$, denoted by $\rho(\varphi,\signal{},t)$, represents the smallest difference between the actual signal value and the threshold at which the system violates $\varphi$.
For example, if the property $\varphi$ says that ``the drone should maintain an altitude of at least 5.0 meters," then $\rho(\varphi,\signal{},t)$ represents how close to 5.0 meters the drone maintains its altitude.
Formally, robustness is defined over STL formulas as follows:
\begin{align}
    \rho(f(\signal{}(t)) > 0,\signal{},t) & \equiv f(\signal{}(t))\nonumber\\
    \rho(\neg \varphi, \signal{}, t) &\equiv -\rho(\varphi,\signal{},t) \nonumber \\
    \rho(\varphi_1\wedge \varphi_2,s,t) &\equiv \min\{\rho(\varphi_1,\signal{},t),\rho(\varphi_2,\signal{},t)\}\nonumber\\
    \rho(\Diamond_{[a,b]}\varphi,\signal{},t) &\equiv \sup_{t_1\in[t+a,t+b]} \rho(\varphi,\signal{},t_1) \nonumber\\
    \rho(\Box_{[a,b]}\varphi,\signal{},t) &\equiv \inf_{t_1\in[t+a,t+b]} \rho(\varphi,\signal{},t_1)\nonumber 
\end{align}
where $\inf_{x \in X} f(x)$ is the greatest lower bound of some function $f:X\rightarrow \mathbb{R}$ (and $\sup$ the least upper bound).
The robustness of satisfying predicate $f(\signal{}(t)) > 0$ captures how close signal $\signal{}$ at time $t$ is above zero.

\paragraph{Parametric signal temporal logic (PSTL)}

PSTL~\cite{PSTL} is a logic obtained by replacing the constants in an STL formula with parameters. The syntax of PSTL is as follows:
\begin{align*}
    \varphi ~:=~    & f(\signal(t)) > a                       ~\mid~
                    \neg \varphi                    ~\mid~
                    \varphi \land \psi       ~\mid~
                    \varphi ~\mathcal{U}_I~ \psi
\end{align*}
Note that the syntax is similar to that of STL except both $a$ and the time interval $I$ can either be a parameter or a constant. In addition, there are two types of parameters in PSTL formulas: $a$ represents the value parameter for the atomic proposition $f(\signal{}(t)) > a$,
whereas $I$ represents the time parameters $[\tau_1, \tau_2]$ (where $\tau_1 < \tau_2$). A PSTL formula is denoted as $\varphi(\parameter{})$, where $\parameter{} = (p_1, ..., p_m) \in \mathcal{P}$ is the tuple of parameters appearing in the PSTL formula.

To instantiate a PSTL formula into an STL formula, a \emph{valuation function} $\nu$ is needed to map parameters to their corresponding concrete values. For example, it maps value parameters to the signal domain $D$, and time parameters to the time domain $T$. A PSTL formula combined with a valuation function $\nu$ for $\parameter$ defines an STL formula $\varphi(\nu(\parameter))$. We say that a PSTL formula $\varphi$ is satisfiable with respect to signal trace $\signal{}$ if there exists an instantiated STL formula $\varphi(\nu(\parameter))$ such that it is satisfiable. It is formally denoted as follows:
\begin{align}
    (\signal{}, t) \models \varphi \Leftrightarrow \exists ~\varphi(\nu(\parameter)) \bullet (\signal{}, t) \models \varphi(\nu(\parameter))
\end{align}
For example, consider PSTL formula $\eventually_{[\tau_1, \tau_2]}(f(\signal(t)) > a)$. Instantiating it with the valuation function $\nu = \{\tau_1 \mapsto 0, \tau_2 \mapsto 5, a \mapsto 30\}$ results in the STL formula $\varphi(\nu(\parameter))$ $\equiv$ $\eventually_{[0, 5]}(f(\signal(t) > 30)$. Given signal $\signal{}(t) = (20,30,40 ...)$,  PSTL formula $\varphi$ is satisfiable with respect to signal $\signal{}$ because the instantiation $\varphi(\nu(\parameter))$ is satisfiable with respect to  $\signal{}$.

\paragraph{Validity Domain} The  validity domain of a PSTL formula is the set of all the parameter values that yield satisfaction for an arbitrary signal \signal{}. We will be extending this concept in Section \ref{section:requirement-adaptation}  for our adaptation framework.




\section{Runtime Adaptation Architecture}
\label{section:runtime-adaptation-architecture}
\begin{figure}[h]
\centering
\includegraphics[width=0.50\textwidth]{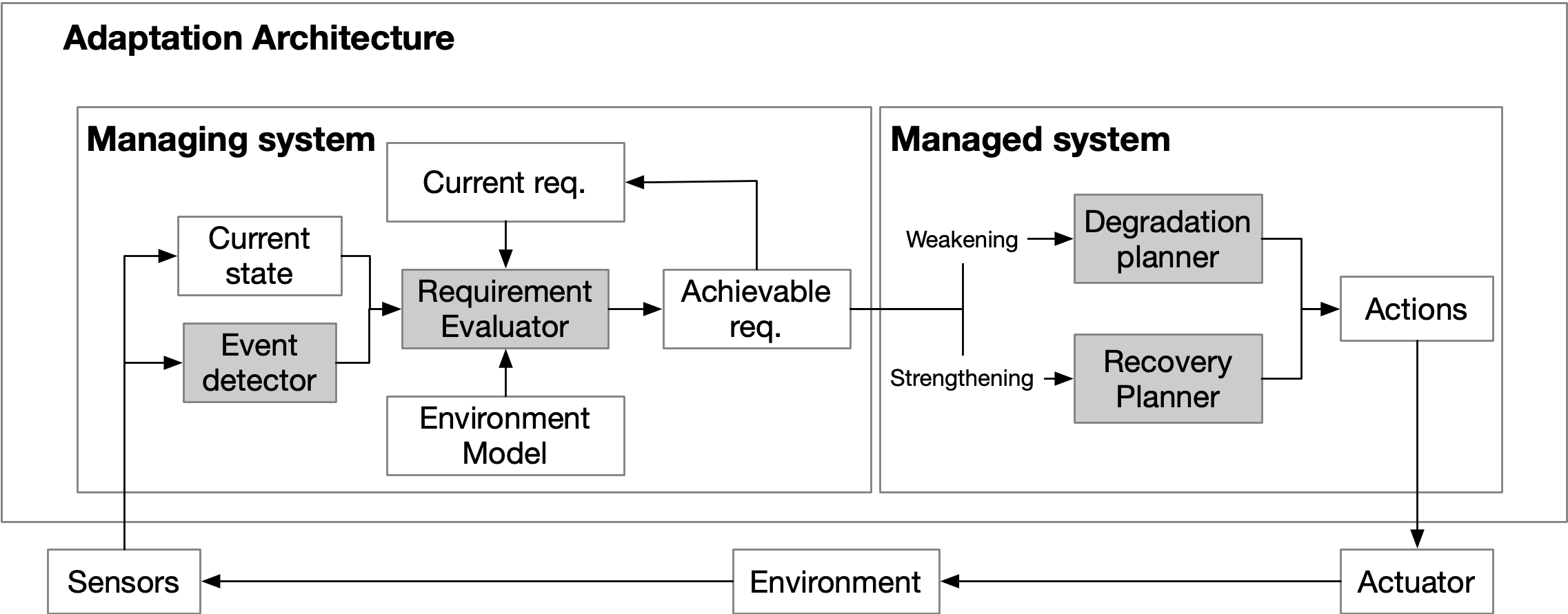}
\caption{Proposed adaptation framework}
\label{figure:runtime-recovery-framework}
\end{figure}

We present an overview of our proposed runtime adaptation architecture in Figure \ref{figure:runtime-recovery-framework}. The adaptation framework periodically observes the state of the environment, determines whether adaptation is needed, and generates actions based on requirements to affect the system and environmental states.



There are three major components in the proposed adaptation architecture. First, the \emph{event detector} looks for degradation or restoration events in the environment. Second, the \emph{requirement evaluator} determines the achievable requirement based on the current environmental conditions. Third, the \emph{degradation} and \emph{recovery planner} plan future system actions based on the changing requirements. For example, if visibility decreases below a certain threshold and the weakened requirement states that "regain contact within the next 5 seconds",
the degradation planner may institute a \emph{wait} action instead of diving directly or diving with a lower speed.

When activated by the \textit{event detector}, the \textit{requirement evaluator} takes various inputs: (1) an environmental event, (2) the current system requirement, (3) the current state of the system, and (4) the \emph{environmental model}, which captures how the state of the environment changes based on system actions. Using a \emph{model-predictive control (MPC)} approach~\cite{mpc}, the evaluator finds an optimal requirement either via strengthening or weakening of the current requirement and produces the corresponding control actions through the planner. 

In the following, we describe (A) the environmental model in more detail, (B) how degradation or adaptation is triggered, and (C) building on these, a precise formulation of the requirements-driven adaptation problem. 


%


\subsection{Environment Model}


Given the current requirement $\varphi_0$, the goal of the evaluator is to search for an alternative requirement $\varphi_1$ that is satisfiable. Evaluating the satisfiability of an STL formula requires knowledge of certain future steps \signal{}, and the environmental model enables the generation of the predictive signals for these particular evaluations.

Formally, the environment model is represented as transition system $T = (\mathcal{Q}, \mathcal{A}, \mathcal{\delta}, \mathcal{Q}_i)$, where:
\begin{itemize}
    \item $\mathcal{Q} \subseteq \mathbb{R}^{k}$ is the set of environment states. Each state is a  combination of values for signal variables, represented as a k-dimensional tuple;  $q = (v_1, ... v_k) \in \mathcal{Q}$.
    \item $\mathcal{A}$ is the set of all  actions, $\mathcal{A} = \mathcal{A}_{sys} \cup \mathcal{A}_{env}$, where $\mathcal{A}_{sys}$ is the set of actuator actions and $\mathcal{A}_{env}$ is the set of environmental actions. Note that $\mathcal{A}_{sys}$ and $\mathcal{A}_{env}$ are disjoint, meaning that $\mathcal{A}_{sys} \cap \mathcal{A}_{env} = \emptyset$.
    \item $\mathcal{\delta}: \mathcal{Q} \times \mathcal{A} \rightarrow \mathcal{Q}$ is the transition function that captures how the system moves from one state to another by performing an action.
    \item $\mathcal{Q}_i$ is the set of initial states.
\end{itemize}

For example, the environment model for the pipeline inspection case study captures the current location and the velocity of the vehicle, the relative location for the pipeline, the thrust of the engine, and how the engine configuration affects the thrust. The location of the vehicle changes during each transition depending on the current velocity, which, in turn, may be modified by a system action that accelerates or decelerates the vehicle (represented by a velocity vector). Suppose the state $q$ is represented as a tuple $(vel_x, vel_y, vel_z)$. The next state can be computed using the previous state $q$, action $a$, and transition function $\delta$, such that $q' = \delta(q, a)$ The example below captures the environmental model for setting the velocity of the vehicle based on the acceleration provided by the engine thrust:
\begin{align*}
q'.vel = (q.vel_x + q.acc_x, q.vel_y + q.acc_y, q.vel_z + q.acc_z)  
\end{align*}
Then, given a sequence of actions $a_1,a_2,...,a_n$ and  current state $q$, the environment model can be executed over these actions to generate a corresponding state sequence, $q_0, q_1, ... q_n$, which is then  formed into predictive signal \signal{}.

We do not impose restrictions on one particular notation for specifying an environment model, as long as it can be used to generate signals in the format illustrated above. With a more powerful backend like Gurobi \cite{qp-gurobi}, one can encode more complex environmental models like non-linear dynamics. For our implementation, we use the MiniZinc modeling language~\cite{minizinc}, which provides declarative constraints for specifying relationships between different variables of a system.

\subsection{Adaptation Trigger}

A key decision in our adaptation process is determining when degradation or recovery should be triggered. Degradation takes place when the system is no longer capable of satisfying the given requirement $\varphi$ in the current environment. We state this condition more formally as follows:
\begin{align*}
& \exists \actionseq_{env} : Seq(\mathcal{A}_{env}) \bullet \forall \actionseq_{sys} : Seq(\mathcal{A}_{sys}) \bullet\\
& \quad \quad  \forall \signal : \Signal \bullet \signal{} = \delta^{*}(q_0, \actionseq_{env} \oplus \actionseq_{sys}) \implies (\signal{},0) \not \models \varphi
\label{degradation-condition}
\end{align*}
where $\Signal$ is the set of all signals, $q_0$ is the current state of the system, $\oplus$ is an interleaving of two sequences of actions. In other words, the above statement says that the environment can behave in a certain way (through some sequence of actions, $\actionseq_{env}$) such that  no matter how the system responds, it is unable to satisfy the  property $\varphi$. The idea is then to carry out degradation to find and satisfy a weaker version of $\varphi$.

Similarly, the triggering condition for recovery is stated as follows:
\begin{align*}
& \forall \actionseq_{env} : Seq(\mathcal{A}_{env}) \bullet \exists \actionseq_{sys} : Seq(\mathcal{A}_{sys}) \bullet\\
& \quad \quad \forall \signal: \Signal \bullet \signal = \delta^{*}(q_0, \actionseq_{env} \oplus \actionseq_{sys}) \land (\signal,0) \models \varphi
\end{align*}
In other words, in the current environment, the system can guarantee the satisfaction of $\varphi$, no matter how the environment behaves. 
Since the environment is in such an agreeable condition, the system may then attempt to improve its utility by satisfying a stronger version of $\varphi$.




Evaluating the above conditions involves generating a predictive signal ($\signal$) that describes how the environment evolves given a sequence of system actions. However, carrying this out frequently may incur significant runtime overhead and possibly interfere with the system operations. Thus, instead of evaluating these conditions, our \emph{event detector} looks for designated \emph{degradation} and \emph{restoration} events ($A_{degrade}$ and $A_{restore}$) and uses these as proxy triggers for degradation and recovery, respectively. Degradation events are abnormal events that occur due to an unexpected change in the environment, such as a thruster failure or unusually low water visibility. On the other hand, a restoration event indicates the environment returning to its previous state, such as an improvement in visibility or recovery of an engine thruster.

\subsection{Adaptation Problems}


We provide a precise statement of our requirements-driven adaptation problems:


\begin{problem}
\textbf{Graceful Degradation Problem}.
\label{prob:degradation}
Given degradation event $a_{0} \in \mathcal{A}_{degrade}$ and current requirement $\varphi_{curr}$, find $\varphi_{curr}'$ and action sequence $\actionseq$ such that
{
\begin{align}
&\forall \signal{} : \Signal \bullet (\signal{},0) \models \varphi_{curr} \Rightarrow (\signal{},0) \models \varphi_{curr}'~\land  \\
& \exists \signal{}_{pred} : \Signal \cdot (\signal_{curr}~^{\frown} \signal{}_{pred}, 0)   \models \varphi_{curr}' ~\land \\
& \qquad \signal{}_{pred}  = \delta^{*}(q_0, \langle a_{0} \rangle ~^{\frown} \actionseq) 
\end{align}
}
\end{problem}
where $q_0$ is the current state of the system (encoded in $\signal_{curr}$) and  $^{\frown}$ is the concatenation operator that is used to link two sequences together (i.e., $\langle s_1, s_2 \rangle ^{\frown} \langle s_3, s_4\rangle$ results in $\langle s_1, s_2, s_3, s_4 \rangle$). Degradation actions $\mathcal{A}_{degrade}$ is a subset of all environmental actions $\mathcal{A}_{env}$ (i.e., $\mathcal{A}_{degrade} \subseteq \mathcal{A}_{env}$) defined in Section \ref{section:runtime-adaptation-architecture}. $\signal{}_{pred}$ represents the future signal generated by the sequence of  actions $\actionseq$, and $\signal{}_{curr}$ represents the signal that encompasses the current state, which is monitored by the sensor. Note also that the new requirement needs to be weaker than the current requirement to achieve degradation, using the definition in Eq. (2), which will be formally defined in Section \ref{section:requirement-adaptation}.


Informally, this problem involves given a degradation event, how to find control actions that can satisfy an alternative requirement such that it still provides an acceptable level of system utility. 

\begin{problem}
\label{prob:recovery} \textbf{Recovery Problem}. Given restoration event $a_{0} \in \mathcal{A}_{restore}$ and current requirement $\varphi_{curr}$, find $\varphi_{curr}''$ and $\actionseq$ such that
{
\begin{align}
&\forall \signal{} : \Signal \bullet (\signal{},0) \models \varphi_{curr}''  \Rightarrow (\signal{},0) \models \varphi_{curr} ~\land 
\label{eq:strengthening} \\ 
& \exists \signal{}_{pred} : \Signal \cdot (\signal_{curr}~ ^{\frown} \signal{}_{pred}, 0)   \models \varphi_{curr}'' ~\land \\
& \qquad \signal{}_{pred}  = \delta^{*}(q_0, \langle a_{0} \rangle ~^{\frown} \actionseq)  
\end{align}
}
\end{problem}


Note that in the case of recovery, the new requirement needs to be stronger than the current requirement to achieve recovery, as defined in Eq. (\ref{eq:strengthening}). The formalism for strengthening will be formally defined in Section \ref{section:requirement-adaptation}.
Informally, the recovery problem is framed as given a restoration event, how to find control actions that can satisfy an alternative requirement such that it can provide an improved level of system utility. 

\section{Requirement Weakening and Strengthening}
\label{section:requirement-adaptation}

In this section, we present an extension to PSTL to (1) incorporate the concept of requirement weakening and strengthening, and (2) restrict the search space of alternative requirements by providing upper and lower bounds for the validity domain of PSTL formulas. We also propose metrics to compare multiple PSTL instantiations. These metrics are then used by the MILP solver to find an optimal requirement; i.e., a minimally weakened or maximally strengthened version of the current requirement.

\subsection{Minimal, Optimal and Current requirements}
\label{paragraph:ref-opt-min-req}
To enable requirement adaptation, we introduce three new concepts that guide and restrict the range of the requirement space. The \emph{minimal requirement}, $\varphi_{min}$, represents the lower bound of the PSTL requirement, allowing for the loosening of constraints when necessary for system adaptability. Conversely, the optimal requirement, $\varphi_{opt}$, signifies the upper bound of the PSTL requirement, enabling the strengthening of the reference requirement. We assume that any requirements that are stronger than the optimal requirement do not provide additional utility for the system, and on the contrary, any requirement weaker than the minimal requirement will result in behavior that is deemed unacceptable to  stakeholders.

The current requirement, denoted as $\varphi_0$, represents the requirement that the system is trying to achieve. Note that the current requirement should always be weaker than (or equal to) the optimal requirement, while always being stronger than (or equal to) the minimal requirement.


Recall the \emph{visibility} requirement for the UUV example in Section \ref{section:motivating-example}. The optimal requirement, in this case, is $\varphi_{opt}$: \textit{Every time the visual contact with the submarine is lost, regain the visual contact within the next 5 seconds}, formally denoted as $\texttt{visibility} < 20 \Rightarrow \Diamond_{[0,5]}(\texttt{distance\_to\_pipe} < 10)$, assuming that visibility of 20 meters is the threshold that determines whether visual contact is maintained.

Suppose that the UUV designer is willing to accept a weakening of the time interval to regain visual contact to 15 seconds; any time above that threshold would be considered unacceptable. Thus, the requirement $\texttt{visibility} < 20 \Rightarrow \Diamond_{[0,15]}(\texttt{distance\_to\_pipe} < 10)$ is designated as a minimal requirement, and any time in between can be set as the current requirement (for example,  regaining visual contact within 7 seconds).




\subsection{Extension to PSTL}
\paragraph{Strengthening and weakening}
Suppose $\varphi_1$ and $\varphi_2$ are two distinct instantiations of the PSTL formula $\varphi$, such that $\varphi_1 = \varphi(\nu_1(\parameter))$, $\varphi_2 = \varphi(\nu_2(\parameter))$, and $\varphi_1 \neq \varphi_2$. We say that $\varphi_1$ is weaker than $\varphi_2$ if and only if Eq. \ref{eq:weak} is true, denoted as $\varphi_1 \weakerthan \varphi_2$; conversely, $\varphi_1$ is stronger than $\varphi_2$ if and only if Eq. \ref{eq:strong} holds, denoted as $\varphi_1 \strongerthan \varphi_2$.
\begin{align}
        \forall \signal \in T \rightarrow D \cdot (\signal{}, 0) \models \varphi_1 \Rightarrow (\signal{},0) \models \varphi_2
    \label{eq:strong} \\
        \forall \signal \in T \rightarrow D \cdot (\signal{}, 0) \models \varphi_2 \Rightarrow (\signal{},0) \models \varphi_1
        \label{eq:weak}
\end{align}

Suppose we are given a PSTL formula $\varphi$, its reference requirement $\varphi_0$, minimal requirement $\varphi_{min}$ and optimal requirement $\varphi_{opt}$. We define \textit{strong valuation} as a set of valuation functions $\nu_s$ such that $\varphi_0 \weakerthan \varphi(\nu_s(\parameter)) \weakerthan \varphi_{opt}$; and \textit{weak valuation} as a set of valuation functions $\nu_w$, such that $\varphi_{min} \weakerthan \varphi(\nu_w(\parameter)) \weakerthan \varphi_0$. The set of all instantiated STL formulas as a result of a strong valuation $\varphi(\nu_s(\parameter))$ are referred to as \emph{strengthened formulas}; and all instantiated STL formulas as a result of a weak valuation $\varphi(\nu_w(\parameter))$ are referred to as \emph{weakened formulas}.

\paragraph{Bounded Validity Domain}
\label{paragraph:bounded-domain-validity}
The \emph{validity domain} of a PSTL formula~\cite{PSTL}, evaluated with respect to signal trace $\signal{}$, is the set of all the parameter values that yield satisfaction for the trace $\signal{}$. We extend the validity domain concept so that it is bounded by minimal and optimal requirements that we defined above in Section \ref{paragraph:ref-opt-min-req}. 

We first define the $\preceq$ operator: $\textbf{v} \preceq \textbf{v}'$  if and only if $\forall j \bullet \textbf{v}_j \leq \textbf{v}'_j$. The $\oplus$ operator in this context is used for time interval concatenation (i.e., $t \oplus [t_1, t_2]$ is equivalent to $[t + t_1, t + t_2]$). 
Additionally, $\valuation{}_{min}$ and $\valuation{}_{opt}$ correspond to the valuations that instantiate the minimal and optimal requirements $\varphi_{min}$ and $\varphi_{opt}$, respectively. Then, the validity domain of PSTL formula $\varphi$ with respect to a signal $\signal{}$, bounded by $\valuation{}_{min}$ and $\valuation{}_{opt}$,  is denoted as $d(\signal{}, \varphi)$ and defined in the following way: 
\begin{align*}
 & d(\signal{}, f(\signal{}(t)) > a) & = & \{(t, \textbf{v}) : f(\signal{}(t)) < a_\textbf{v} ~\land~ \\
    &&& (\valuation_{min} \preceq \valuation{} \preceq \valuation_{opt}) \} \\
 & d(\signal{}, \varphi \land \psi) & = & d(\signal{}, \varphi) \cap d(\signal{}, \psi) \\
 & d(\signal{}, \neg \varphi) & = & d(\signal{}, \varphi) \\
 & d(\signal{}, \varphi ~\until{}_I~ \psi) & = & \{(t, \textbf{v}) : \exists t' \in t \oplus I_{\textbf{v}} \bullet (t', \textbf{v}) \in d(\signal{}, \psi) \\
    &&& \land \forall t'' \in [t, t'](t'', \textbf{v}) \in d(\signal{}, \varphi) \\
    &&&\land~ (\valuation_{min} \preceq \valuation{} \preceq \valuation_{opt})\} 
\end{align*}
Note that $a_\textbf{v}$ and $I_\textbf{v}$ here either represent constants or the concrete value assignment of the parameter $a$ and $I$ using valuation tuple \textbf{v}. The resulting set, $d(\signal, \varphi)$, is a set of all 2-element tuples of form $(t, \nu(\parameter))$, where $(\signal, t) \models \varphi(\nu(\parameter))$. 



\paragraph{Degree of Weakening and Strengthening}
To quantitatively measure and compare the relative restrictiveness between two instantiated formulas (i.e., $\varphi(\nu_1(\parameter))$ and $\varphi(\nu_2(\parameter))$, which will be abbreviated to $\varphi_1$ and $\varphi_2$ below), we introduce two metrics: \textit{degree of weakening} (Eq. \ref{degree-of-weakening}), and its inverse metric, \textit{degree of strengthening} (Eq. \ref{degree-of-strengthening}). These metrics are defined based on the robustness of satisfaction, as follows:
\begin{align}
\begin{split}
\Delta_{weak}(\varphi_1, \varphi_2, \signal{}, t) & = \rho(\varphi_2, \signal{}, t) - \rho(\varphi_1, \signal{}, t)
\end{split}
\label{degree-of-weakening} \\
\begin{split}
\Delta_{strong}(\varphi_1, \varphi_2, \signal{}, t) & = \rho(\varphi_1, \signal{}, t) -\rho(\varphi_2, \signal{}, t)
\end{split}
\label{degree-of-strengthening}
\end{align}
Note that for \textit{degree of weakening} to be a positive number, $\varphi_2$ must be weaker than $\varphi_1$, and vice versa for \textit{degree of strengthening}. We will present an example below to illustrate how these metrics are used.

\paragraph{Example}
Recall the example in Section \ref{section:motivating-example}, with the requirement for the \textit{thruster} feature: The thruster should provide 100N of thrust within the next second. The parameter 100N is subject to change. This requirement can be formalized in PSTL as $\varphi \equiv \globally_{[0,1]}(\texttt{thrust} > p_1)$. The original requirement is  $\varphi_{origin} \equiv \varphi(p_1 \mapsto 100) \equiv  \globally_{[0,1]}(\texttt{thrust} > 100)$, while one possible weakened version is  $\varphi_{weak} \equiv \varphi(p_1 \mapsto 70) \equiv  \globally_{[0,1]}(\texttt{thrust} > 70)$.

Suppose the system evolves to generate signal \signal{} with a thrust of $110$ and $80$, at time $0$ and $1$ second, respectively; then $\rho(\varphi_{origin}, \signal{}, 0) = -20$ (meaning $\varphi_{origin}$ is violated), while 
$\rho(\varphi_{weak}, \signal{}, 0) = 10$ (meaning $\varphi_{weak}$ is satisfied. Thus, the \textit{degree of weakening} is measured as $\Delta_{weak}(\varphi_{origin}, \varphi_{weak}, \signal{}, 0) = 30$ (meaning requirement weakening from $\varphi_{weak}$ to $\varphi_{origin}$), while \textit{degree of strengthening} is measured as $\Delta_{strong}(\varphi_{weak}, \varphi_{origin}, \signal{}, 0) = 30$ (meaning strengthening from $\varphi_{weak}$ to $\varphi_{origin}$). 



\subsection{Runtime Adaptation as a MILP Formulation}


The requirement adaptation process can be performed by reducing it to an instance of MILP, where the problem consists of solving for a set of decision variables that optimize certain objectives while fulfilling a set of hard constraints. We next show how minimizing the degree of weakening and maximizing the degree of strengthening can be formulated as MLIP objectives.

\begin{problem}
\label{prob:req-exploration-weakening-milp}
\textbf{Graceful Degradation as MILP}.
Given the current requirement $\varphi(\nu(\parameter))$ and its weaker version to be found, $\varphi(\nu'(\parameter))$, transition system $T = (\mathcal{Q},\mathcal{A},\delta,\mathcal{Q}_i)$, and state sequence $q^t = q_0,\dots,q_t$ representing the signal observed from the environment so far, compute: 
\begin{align}
    \text{argmin}_{\nu'(\parameter), \actionseq} ~&~ \Delta_{weak}(\varphi(\nu(\parameter)), \varphi(\nu'(\parameter)), \signal{},0) \label{eq:opt-objective} \\
    s.t. 
    ~&~ \signal{}_i = q_i\ \text{for } i \leq t  \label{eq:opt-past}~\land \\
    ~&~ \signal{}_i = \delta(\signal_{i-1}, \mathbf{a}_{i-1})\ \text{for } t<i\leq t+N\label{eq:opt-future}~\land \\ 
    ~&~ (0, \nu'(\parameter)) \in d(\signal{}, \varphi)
    \label{formula:solver-domain-validity}
\end{align}
\end{problem}
There are two decision variables to be solved in this MILP problem: parameters $\nu'(\parameter)$ and control action sequence $\actionseq$. Note that in Eq. (\ref{eq:opt-past}), the past signal is defined up until the current time. Eq. (\ref{eq:opt-future}) involves the prediction of a future signal from the next time step to time $t + N$ using the transition system $T$ that encodes an environment model. $N \in \mathbb{N}$ is a finite predictive horizon inferred from the STL requirement $\varphi$. Eq. (\ref{formula:solver-domain-validity}) guarantees that the new requirement is within the defined range of requirements by restricting the parameters of the PSTL formula to the validity domain.

Finally, by minimizing the degree of weakening from  $\varphi(\nu(\parameter))$ to $\varphi(\nu'(\parameter))$ in Eq. (\ref{eq:opt-objective}), we guarantee that the system utility associated with the requirement is degraded by a minimal necessary amount.


The formulation of recovery as MILP mirrors that for degradation:
\begin{problem}
\label{prob:req-exploration-strengthening-milp}
\textbf{Recovery MILP Formulation}.
Given the current requirement $\varphi(\nu(\parameter))$ and it strengthened version to be found, $\varphi(\nu'(\parameter))$, transition system $T = (\mathcal{Q},\mathcal{A},\delta,\mathcal{Q}_i)$, and state sequence $q^t = q_0,\dots,q_t$ representing the signal observed from the environment so far, compute: 
\begin{align}
    \text{argmax}_{\nu'(\parameter), \mathbf{a}} ~&~ \Delta_{strong}(\varphi(\nu(\parameter)), \varphi(\nu'(\parameter)), \signal{},0) \label{recovery-eq:opt-objective} \\
    s.t. 
    ~&~ \signal{}_i = q_i\ \text{for } i \leq t  \label{recovery-eq:opt-past}~\land \\
    ~&~ \signal{}_i = \delta(\signal_{i-1}, \mathbf{a}_{i-1})\ \text{for } t<i\leq t+N\label{recovery-eq:opt-future}~\land \\ 
    ~&~ (0, \nu'(\parameter)) \in d(\signal{}, \varphi)
    \label{recovery-formula:solver-domain-validity}
\end{align}
\end{problem}
In comparison to the MILP formulation for the degradation problem, the objective here is to  \emph{maximize} the degree of strengthening between $\varphi(\nu(\parameter))$ and $\varphi(\nu'(\parameter))$, as shown in Eq. (\ref{recovery-eq:opt-objective}). This allows searching for the best possible requirement that allows the system to regain utility that was temporarily sacrificed during an earlier degradation step.

\section{Implementation}

\subsection{Simulator}
\begin{figure}[t]
\centering
\includegraphics[width=0.3\textwidth]{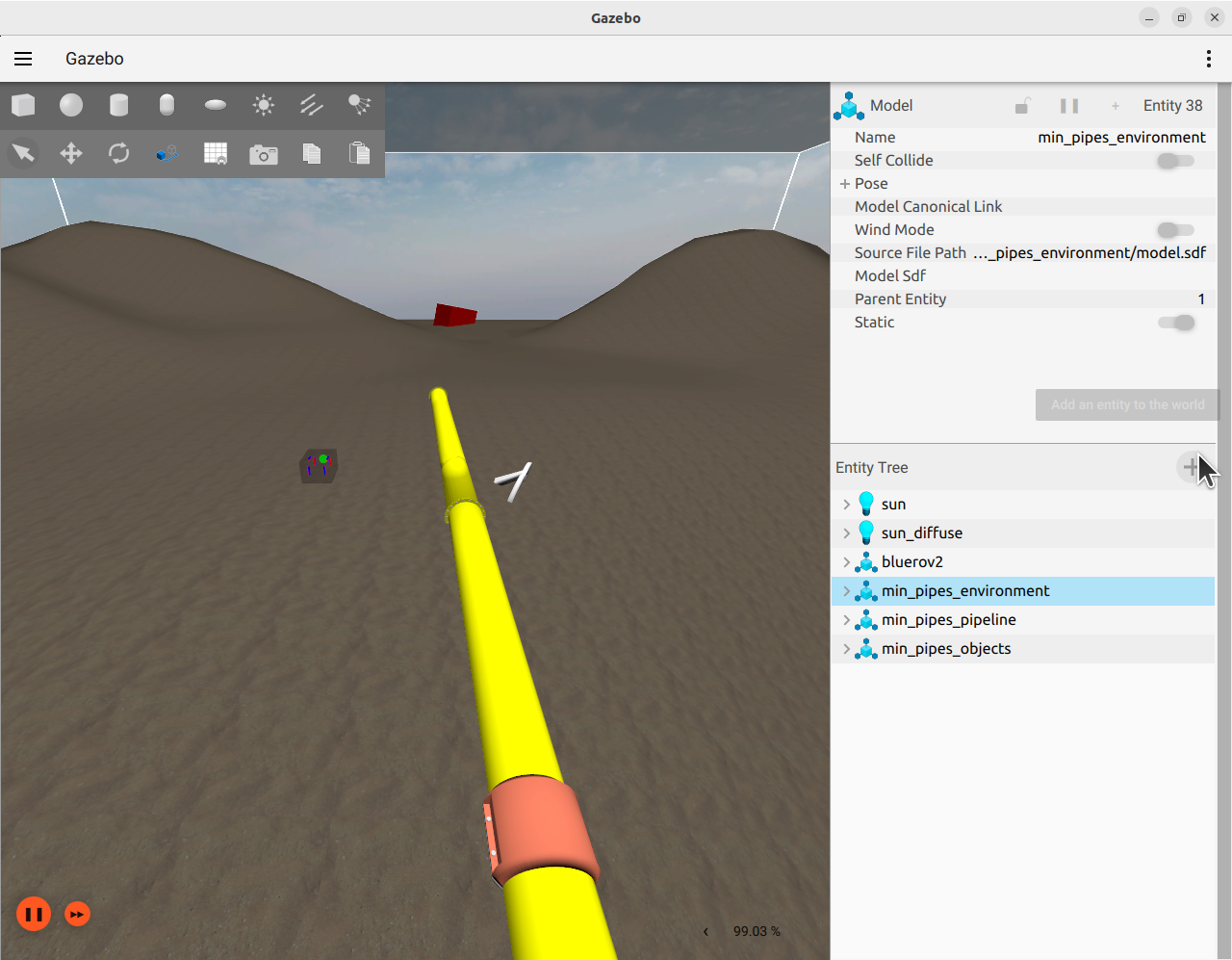}
\caption{A screenshot of the Gazebo UUV simulator.}
\label{fig:gazebo-simulator}
\end{figure}

To illustrate our proposed requirement adaptation approach, we have developed a prototype implementation\footnote{All the code, models, and experimental data are available at https://github.com/sychoo/cps-degradation-recovery} based on SUAVE \cite{suave-case-study}, an unmanned underwater vehicle that supports the customization of self-adaptation logic. It uses a ROS2-based platform and implements a pre-defined mission that detects, follows, and inspects a pipeline on the seabed. The backend of the simulator uses ArduSub \cite{ardusub-simulator}, which provides various controller APIs to control the trajectory of the underwater vehicle. The vehicle trajectory and environmental setup can be visualized through the Gazebo simulator \cite{ros-gazebo} with the UI shown in Fig. \ref{fig:gazebo-simulator} where the seabed pipe is indicated with the yellow cylinder, while the UUV is illustrated by the rectangular box to the left side of the pipe. Various environmental entities (i.e., lighting, terrain) are listed on the panel to the right, and can be adjusted as needed.






\textbf{Features.} For evaluation, we implemented two mission-related features in the UUV. When these features are activated, they generate propulsion actions (in the form of velocity vectors) or system reconfiguration actions that override those from the path planner. The implemented features are as follows:
\begin{itemize}
    \item \textit{Visibility Monitor} ensures that the UUV maintains sufficient visibility to the pipelines. To achieve this, it generates actions to enforce the UUV to close in on the pipelines when the visibility is below a safety threshold.
    
    \item \textit{Thrust Monitor} ensures that the UUV maintains enough thrust to support the timely completion of the mission. The thrust monitor may change the system configuration dynamically (i.e., turning on additional thrusters when the actual thrust falls below the expected thrust.)
\end{itemize}

\textbf{Environmental Anomalies.} Next, we randomly injected abnormal environmental events throughout the operation of the UUV. We list two failures that we investigated:

\begin{itemize}
    \item \textit{Loss of visual contact with the underwater pipeline}: The low water visibility makes it difficult for the UUV to detect and follow the underwater pipeline. When it occurs during the  pipeline inspection, it may result in the UUV losing sight of the pipeline and requiring it to dive deeper in the short term, regain the visual line of sight, and continue the inspection progress. 
   The change in the water visibility is introduced randomly during the simulation. 
    \item \textit{Thruster failure}: A thruster failure causes the engine of the UUV to provide partial propulsion, move erratically, or shut down completely. Similarly, thruster failures are modeled and injected in a stochastic manner.

\end{itemize}

\subsection{Environment Model}
Our framework leverages a model of the environment during the adaptation process. For the UUV system, the environmental model captures the (1) physical dynamics of the system (2) thruster estimation from the engine, and (3) the estimated coordinates of the seabed pipeline. The dynamics model is used to estimate the velocity of the UUV based on the accelerating or decelerating actions. The thruster estimation is based on the configuration of the thrusters (i.e., which ones are turned on or off). Lastly, the estimated coordinates of the seabed pipeline are based on information received via onboard sensors. The coordinate information is also used to keep track of the inspection progress and guide the diving operation toward the pipeline.

All aspects of the environmental model are specified in the MiniZinc modeling language, translated to linear constraints and solved using MILP during the  adaptation process.

\section{Evaluation}



This section describes the evaluation of our proposed approach. We present the following:
\begin{itemize}
    \item \textbf{RQ1}: Does our approach achieve a higher overall system utility than existing state-of-the-art approaches?
    \item \textbf{RQ2}: What is the runtime overhead of the proposed approach? Does it interfere with system operations?
\end{itemize}


\subsection{Experimental Design}
\label{subsection:evaluation-experimental-design}
We conducted a set of experiments to evaluate the proposed adaptation approach through comparison against with state-of-the-art self-adaptation method in TOMASys~\cite{meta-model}.To ensure that our adaptation approach generalizes across various scenarios, we randomly generated 100 different system setups and failure scenarios, including the duration of the simulation, times when failures are injected, changes in the water visibility, number of usable thrusters, the initial position of the UUV,  and the location of the underwater pipeline.
To compare our proposed method and the baseline approach, we use the \emph{cumulative utility} as the metric, which is measured by calculating the robustness of the minimal requirement for the signal  collected during the degradation and recovery process. The reason why we chose to measure the satisfaction of the minimal requirement is that it is the lower bound of the instantiated STL formula under which the system feature ceases to function and provides useful utility. We assume that the robustness metric is a suitable medium to reflect the desirability of the system behavior. The robustness is measured upon the start of the degradation events and ends upon the satisfaction of the optimal requirement ($\varphi_{opt}$) which indicates the end of the recovery. If the optimal requirement is never reached due to persistent environmental disruptions, the cumulative robustness measurement will continue until the end of the simulation life cycle.



Before we conducted the experiments for the case studies, we created the following hypotheses to be tested: 

\begin{itemize}
    \item \textbf{H1}: Our approach results in a higher cumulative utility throughout degradation and recovery processes than the existing method.
    \item \textbf{H2}: Our  approach results in a higher run-time overhead but it does not disrupt normal system operations. 
\end{itemize}
All our experiments were run on a Ubuntu desktop machine with 16 GB RAM, 6-core Intel Core i5, and a 
NVIDIA GeForce RTX 3060 graphics card.

\subsection{Seabed Pipe Inspection Case Study}


Recall the example mentioned in Section \ref{section:motivating-example}, where a UUV is conducting a mission to inspect underwater pipes on the seabed. The optimal STL formula $\varphi_{opt}$ and the minimal requirement $\varphi_{min}$ for the water visibility and thrust requirement are specified as follows:





\begin{align}
\begin{split}
  & \varphi_{visibility\_opt}: \globally_{[0,1]}(\texttt{visibility} < 20 \Rightarrow \\ & \quad \quad \quad \quad \quad \eventually_{[0,5]}\texttt{distance\_to\_pipeline} < 10)
\end{split}
\label{visibility-opt}\\
\begin{split}
  & \varphi_{visibility\_min}: \globally_{[0,1]}(\texttt{visibility} < 20 \Rightarrow \\ & \quad \quad \quad \quad \quad \eventually_{[0,15]}\texttt{distance\_to\_pipeline} < 10)
\end{split}
\label{visibility-min} \\
\begin{split}
  & \varphi_{thruster\_opt}: \globally_{[0,1]} (\texttt{thrust} > 100) 
\end{split}
\label{thruster-opt}\\
\begin{split}
  & \varphi_{thruster\_min}: \globally_{[0,1]} (\texttt{thrust} > 50)
\end{split}
\label{thruster-min}
\end{align}

The visibility requirement $\varphi_{visibility}$ ensures that the UUV can closely observe the pipelines upon the visibility falling below the safety threshold 20. The thrust requirements allow the system to continuously provide adequate propulsion to ensure a timely mission completion.

\begin{figure}[h]
\centering
\includegraphics[width=0.5\textwidth]{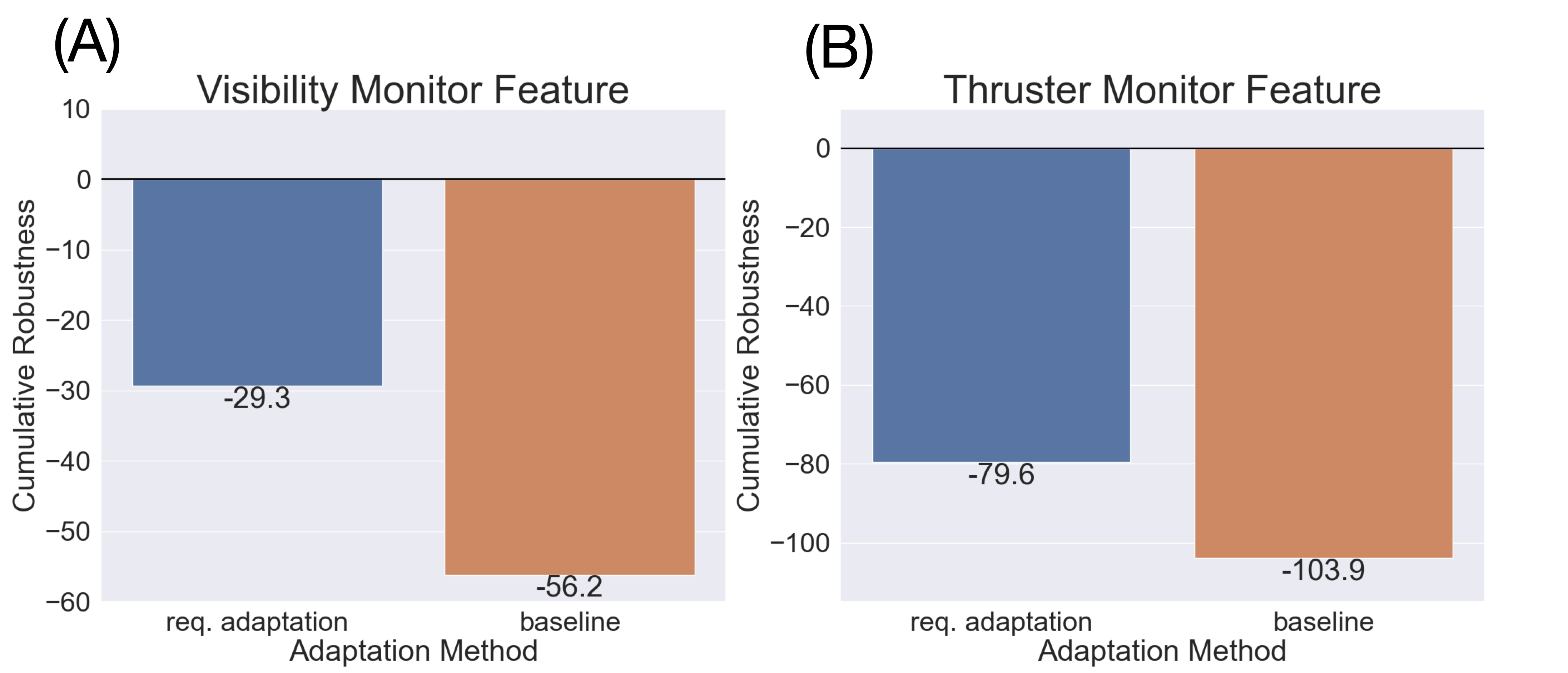}
\caption{Cumulative robustness for visibility and thruster monitor features for the pipeline inspection case study}
\label{fig:overall-result}
\end{figure}

\subsection{Experiment Results}
In Fig \ref{fig:overall-result}, charts (A) and (B) show the cumulative robustness values for both the visibility and thruster monitor feature while the system is encountering thruster failures or low water visibility events. The results show that our adaptation approach achieves a higher cumulative robustness value than the baseline self-adaptive approach. Specifically, in the case of the water visibility monitor, it achieves an approximately 2-fold increase in cumulative robustness (with a 26.9 increase in robustness). In the case of the thruster monitor feature, our approach has a 24.3 lead in cumulative robustness.


The SUAVE artifact \cite{suave-case-study} also comes with a set of mission-related metrics that measure the quality of the mission completion. We reused some of the metrics, namely, the distance of the pipeline inspected. We discovered that our approach, on average, can inspect 86.10 meters of the pipeline while the baseline approach can only inspect 49.63 meters, an increase of about 74\%. However, the standard deviation of our approach (36.68) is slightly worse than the baseline approach (23.30), meaning that the actual performance fluctuates more across various failure scenarios in the 100 experiment runs. 

There are two main reasons for this increase in the cumulative utility and the quality of the mission completion. First, it is the dynamic nature of the action planning based on the specific scenario or context. For instance, TOMASys relies on a fixed set of actions (determined at design time) to address each scenario. To satisfy the objective of regaining contact with the pipeline after losing visuals, TOMASys simply has a predefined action to dive several meters down the seabed, whereas our approach predicts all possible signals given available actions using an environmental model. Then it chooses the actions that result in the most desirable signals. Second, it is the ability to flexibly adapt goals in different situations---namely, adjusting the requirements when they are not attainable or have the potential for improvements. The dynamic action planning and the requirement adjustments result in a more measurable system utility, and therefore, more robust and desirable system actions.








\subsection{Performance Overhead}
Since the requirement adaptation approach requires the use of MILP to search for requirements and plan actions at run time, it has a higher overhead than the baseline approach that uses actions that are defined at design time. We measure the overhead as the average additional time the system uses per control cycle as a result of deploying our adaptation approach compared to the baseline approach using TOMASys. Consequently, we measured the average overhead across both the visibility and thrust monitor features as 0.35 seconds. We have not observed noticeable delays or disruptions to our UUV during the operation, as the UUV controller was running at 2Hz, resulting in a window of 0.5 seconds for each cycle of the controller update.

Furthermore, the performance overhead is subject to the complexity of both the feature requirements and the environmental model. For example, the visibility monitor feature incurs a higher overhead because the encoding of the STL formula results in a more complex set of constraints for the MILP solver at runtime. 

\subsection{Threats to Validity}

\textbf{Internal Validity}: The sampled configuration space is partial and may not capture all exceptional scenarios exhaustively. However, we have mitigated selection biases by randomizing the configuration parameters. For example, at which point failures occur, the initial location of the UUV, the setup of the underwater pipelines, etc. In addition, we are using a deterministic environmental model that captures simple physical dynamics. This is to simplify the scope of the problem and reduce the search space for the changed requirements and corresponding actions. However, the model may not capture the dynamic of the real world caused by external factors (i.e., UUVs may deviate from the direction it is accelerating towards due to water currents). We attempted to mitigate this discrepancy by manually inspecting the source code of the ArduSub simulator to ensure the model largely captures the logic of the simulator.

\textbf{External Validity}: The overhead of the simulation is application-specific and hardware-dependent. More restrictive hardware or outdated MILP solver may increase the runtime overhead of our framework and therefore require more performance tuning than our existing software artifacts. 

Furthermore, the result of the case study may not be generalized to all applications of CPS. Depending on the specific applications (i.e., electricity grids, UAVs), the system requirement may be expressed differently, and so does the way to degrade the systems. However, we believe the concept of using requirement weakening and strengthening to perform degradation and recovery is requirement-agnostic, and that it can be generalized across different domains.

\section{Related Work}


\textbf{Graceful Degradation and Recovery.} Graceful degradation is the ability of a system to maintain an acceptable level of functionality even when a significant portion of the system is rendered inoperable due to environmental disturbances. Recovery refers to the ability of the system to restore to a safe or desired state after a failure. Both areas are well-studied in control systems and human-machine interfaces. Many degradation frameworks have been developed for different applications and domains, such as adaptive cruise control~\cite{graceful-degradation-cruise-control}, autonomous drones~\cite{feature-interaction-weakening,uav-mission-adaptation}, software user interfaces~\cite{graceful-degradation-UI}, industrial control systems~\cite{controller-weakening, automated-graceful-degradation}, and design patterns for degradation\cite{graceful-degradation-design-pattern}, respectively. There is also a large body of work on system recovery. \cite{chen22} proposed an STL-based resilient framework that enables analysis of trade-offs between time-to-recovery and durability in a cruise control system using multi-objective optimization, while \cite{kong18} and \cite{rollback-recovery} propose checkpointing techniques (i.e., storing system traces with a finite horizon leading up to the current state). Lastly, \cite{restart-recovery} and \cite{reset-recovery} use system restart and reset to recover the system from a failure. As far as we know, no prior work provides a coordination mechanism to facilitate these two processes in a single system.

\textbf{Self-adaptive Systems.} Self-adaptive systems aim to design systems that are capable of adjusting to changes in the environment. The most influential reference control model in autonomic and self-adaptive systems is the MAPE-K \cite{mape-k} architecture, which stands for Monitor-Analyze-Plan-Execute over a shared Knowledge, on which our resolution architecture is based. Self-healing systems \cite{shaw02} have been proposed by Shaw, et al. to address the problem of coping with fluctuations and uncertainties in the environment. The idea is that the system always keeps a background maintenance process running regardless of the current state of the system and the environment (i.e. garbage collection, optimized network routing, etc.). This way, the system will eventually maintain internal stability and continuous improvement despite external changes. This state is also known as a homeostasis state. Our approach draws inspiration from this idea, in that we attempt to ensure that the system aims to achieve the most desirable requirement relative to the current requirement. 



The concept of \emph{meta-control} has been used for design-time analysis. For example, OpenODD \cite{open-odd}  has been used as a scenario-based analysis framework for autonomous systems, especially in self-driving cars. TOMASys \cite{meta-model} provides a  comprehensive set of reconfiguration strategies ranging from architectural reconfiguration, to monitor construction, and then to scenario-based definition of adaptation tactics. 


\textbf{Requirement Adaptation.} Requirement adaptation has been investigated in the context of self-adaptive systems. 
The idea of weakening a requirement has been studied for feature interaction resolution~\cite{feature-interaction-weakening}, handling a violation of environmental assumptions and safety properties~\cite{controller-weakening,uav-mission-adaptation}, controller synthesis~\cite{gr1-controller-synthesis,temporal-relaxation}, and goal adjustment for security-critical systems~\cite{weakening-security-critical-system}. As far as we know, however, these existing works focus on how to gracefully degrade the system using requirement relaxation, instead of combining both degradation and recovery stages or attempting to coordinate them using requirement adaptation. 

RELAX \cite{relax} is designed to support expressing requirements that explicitly capture uncertainties about runtime system behavior based on fuzzy temporal logic. For example, with RELAX, users can specify requirements like, "Once a user request is sent, it should be processed as early as possible". Although our approach differs in that it relies on STL as the underlying  specification formalism, an interesting future work would be to investigate the utility of RELAX for the type of degradation-recovery loop that we have explored.




Both \cite{wohlrab-runtime-planning} and \cite{nianyu-preference} study goal adaptation in the context of changing stakeholder goals or user preferences. However, their proposed methods are designed to be dependent on human inputs, not for automatically adapting to changing environments. \cite{resource-driven-adaptation} focuses on requirement adaptation based on resource constraints and proposes a 2-tier adaptation framework that (1) first tries to fulfill goals by finding alternative resources and (2) if needed, deviates the goal slightly from the original one to compensate for limited resources. Our approach differs from these approaches in that  our utility is represented as the desirability of requirements over run-time signals, instead of adequate resources being allocated.

\section{Limitations and Future Work}
We propose a requirement-driven runtime adaptation framework that coordinates between grace degradation and recovery. Through the case study on UUVs, we have demonstrated our proposed approach can result in more desirable system behaviors during periods of degradation and recovery.

However, our approach makes several assumptions about the system that may limit its applicability. First, our approach tackles the class of requirements can be specified using STL \cite{stl}. These requirements are specific to cyber-physical systems, where the system is time-sensitive, and sensor inputs can be monitored and transformed into continuous signals and quantified using state predicates. Thus, the framework is currently not designed to handle other classes of requirements (e.g., discrete behavioral specifications in LTL or  stochastic ones in PCTL\cite{PCTL}).

Secondly, we assume the system has a meaningful way of quantifying the satisfaction of its requirements,  these requirements can be weakened, and the user is willing to tolerate temporary degradation in the associated system utility. Thus, this framework is not well-suited for safety-critical requirements, which tend to be hard constraints and cannot be negotiated (i.e., a collision avoidance system for smart vehicles).

Thirdly, we assume that  interactions between the system and the external environment (i.e. how a UUV move around the water based on an actuator command) can be predicted by a deterministic \textit{environmental model}. However, the estimation from the environmental model may deviate from reality, due to environmental disturbance and other uncertainties. To address this, we used model predictive control \cite{mpc}, which involves repeated prediction and planning on a short horizon to reduce the effect of deviation in the estimation. 

Another limitation of our approach is that a MILP solver may fail to find a satisfiable solution. This can be due to an overly restrictive environmental model, a limited range of alternative requirements that the solver can search for, or an unresolvable system state at the time of the adaptation. While this happened rarely in our case studies (less than 2\% of adaptation cycles), we can cope with the problem through some minor tuning of the adaptation frameworks like reducing constraints to the environmental model, increasing the ranges of requirements available, or defining some simple fallback adaptation approaches, etc.


In future work, we also plan to investigate the applicability of our approach in our domains, such as robotics and cyber-security, where the ability to gracefully degrade and recover is critical for system autonomy and resilience.

\section{Acknowledgement}
This work was  supported in part by Award N00014172899 from the Office of Naval Research, Award No. H98230-23-C-0274 from the National Security Agency, and the National Science Foundation award CCF-2144860. Any views, opinions, findings, conclusions, or recommendations expressed in this material are those of the author(s) and do not necessarily reflect the views of the funding agencies.

\bibliographystyle{ACM-Reference-Format}
\bibliography{references}

\end{document}